\documentclass[osajnl,twocolumn,showpacs,superscriptaddress,10pt]{revtex4-1} 
\usepackage{amsmath,amssymb,graphicx}
\usepackage{graphicx}
\begin{document}

\title{Spectral-phase interferometry for direct electric-field reconstruction applied to seeded extreme-ultraviolet free-electron lasers}

\author{Beno\^{i}t Mahieu}\email{Corresponding author: benoit.mahieu@ensta-paristech.fr}
\affiliation{Laboratoire d'Optique Appliqu\'{e}e, UMR 7639, ENSTA-CNRS-Ecole Polytechnique, Chemin de la Huni\`{e}re, 91761 Palaiseau, France}
\author{David Gauthier}
\affiliation{Sincrotrone Trieste Elettra, S.S.14 ‑ km 163.5 in AREA Science Park, 34149 Basovizza, Italy}
\author{Giovanni De Ninno}
\affiliation{Laboratory of Quantum Optics, University of Nova Gorica, Vipavska 11c, 5270 Ajdov\v{s}\v{c}ina, Slovenia}
\affiliation{Sincrotrone Trieste Elettra, S.S.14 ‑ km 163.5 in AREA Science Park, 34149 Basovizza, Italy}
\author{Hugo Dacasa}
\affiliation{Laboratoire d'Optique Appliqu\'{e}e, UMR 7639, ENSTA-CNRS-Ecole Polytechnique, Chemin de la Huni\`{e}re, 91761 Palaiseau, France}
\author{Magali Lozano}
\affiliation{Laboratoire d'Optique Appliqu\'{e}e, UMR 7639, ENSTA-CNRS-Ecole Polytechnique, Chemin de la Huni\`{e}re, 91761 Palaiseau, France}
\author{Jean-Philippe Rousseau}
\affiliation{Laboratoire d'Optique Appliqu\'{e}e, UMR 7639, ENSTA-CNRS-Ecole Polytechnique, Chemin de la Huni\`{e}re, 91761 Palaiseau, France}
\author{Philippe Zeitoun}
\affiliation{Laboratoire d'Optique Appliqu\'{e}e, UMR 7639, ENSTA-CNRS-Ecole Polytechnique, Chemin de la Huni\`{e}re, 91761 Palaiseau, France}
\author{David Garzella}
\affiliation{Service des Photons Atomes et Mol\'{e}cules, Centre d'Etudes de Saclay, Commissariat \`{a} l'Energie Atomique, B\^{a}timent 522, 91191 Gif-sur-Yvette, France}
\author{Hamed Merdji}
\affiliation{Service des Photons Atomes et Mol\'{e}cules, Centre d'Etudes de Saclay, Commissariat \`{a} l'Energie Atomique, B\^{a}timent 522, 91191 Gif-sur-Yvette, France}

\begin{abstract}
We present a setup for complete characterization of femtosecond pulses generated by seeded free-electron lasers (FEL's) in the extreme-ultraviolet spectral region. Two delayed and spectrally shifted replicas are produced and used for spectral phase interferometry for direct electric field reconstruction (SPIDER). We show that it can be achieved by a simple arrangement of the seed laser.  Temporal shape and phase obtained in FEL simulations are well retrieved by the SPIDER reconstruction, allowing to foresee the implementation of this diagnostic on existing and future sources. This will be a significant step towards an experimental investigation and control of FEL spectral phase.
\end{abstract}


\maketitle 

In recent years, seeded free-electron lasers (FEL's) have demonstrated to be very attractive sources for intense light production, revealing improved shot-to-shot stability, tunability, spatial quality and longitudinal coherence \cite{PRLSRFEL,NaturePhotFERMI,seedingHHG}. Notably, the high-gain harmonic generation (HGHG) scheme \cite{Yu} enables the generation of powerful and ultrashort extreme-ultraviolet (XUV) pulses. In such a configuration, an external source (the seed) interacts with a relativistic electron beam wiggling in a first undulator chain (the modulator). This interaction leads to an energy modulation of the electrons, further transposed to a spatial bunching after the electrons experience an energy-dependent path into a magnetic chicane, called the dispersive section. The bunched electron beam is then injected into a long undulator chain (the radiator). The bunching has a periodicity determined by the seed frequency but also presents significant components at the harmonics of the latter, so that the electron beam can emit coherently at one of the seed harmonics. In the radiator, the light is amplified at the chosen harmonic until the process reaches saturation, due to bunching deterioration.

\begin{figure*}[htbp]
\centerline{\includegraphics[width=2\columnwidth]{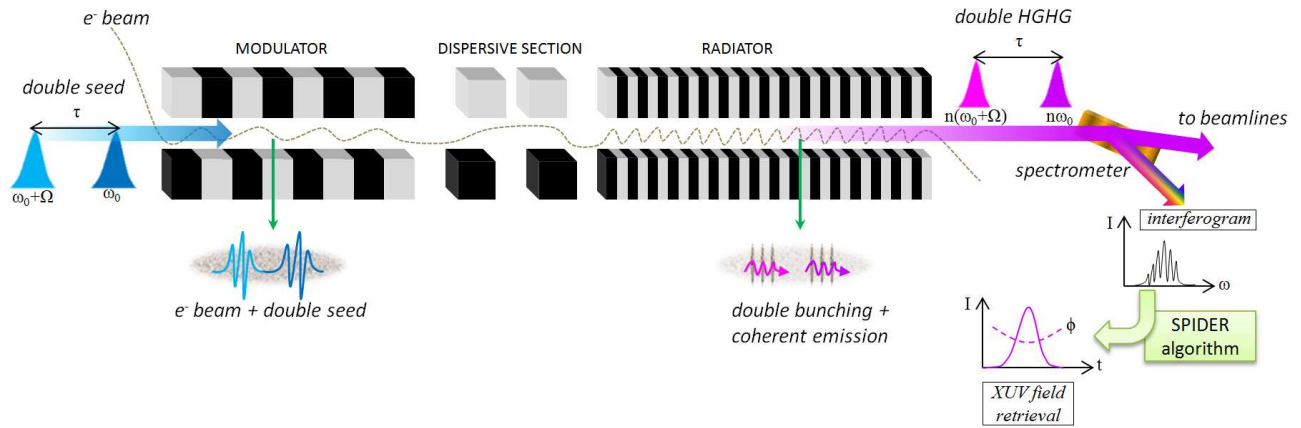}}
\caption{Setup for XUV SPIDER on HGHG FEL. The first seed pulse, of central frequency $\omega_0$, and its replica, delayed by $\tau$ and spectrally shifted by $\Omega$, overlap in space and time with a relativistic electron beam wiggling into the modulator. After the dispersive section, the interaction gives rise to bunching at both seed positions, so that double FEL emission occurs in the radiator, at the harmonic number $n$ of the seed frequencies. The two successive XUV pulses are sent to the spectrometer, which records their interferometric pattern. The latter is analyzed by the direct SPIDER algorithm so that the complete field of the XUV emission is retrieved. In parallel of the measurement, the direct beam (zero-th diffraction order of the spectrometer's grating) is sent to users beamlines.}
\label{FEL}
\end{figure*}

In ideal conditions, the spectro-temporal properties of HGHG mimic those of the seed \cite{PRLSRFEL}, that is usually a Gaussian quasi-monochromatic pulse (e.g., from a Ti:Sapphire laser or an harmonic of the latter).
However, due to the possibility of spectral phase distortion during amplification \cite{Wu}, pure spectral diagnostics does not allow for inferring the temporal pulse shape. Few techniques have been tested on FEL's for directly measuring the temporal profile. They usually rely on photoionization in gaseous targets and remain rather complicated to implement. In their standard layouts, autocorrelation \cite{autoco} and cross-correlation \cite{crossco} require a multi-shot scan (implying source stability). They are thus not suitable for providing, in real time, diagnostic and interconnection with experiments. In addition, autocorrelation cannot resolve pulse asymmetries. More accurate, terahertz streaking spectroscopy \cite{streaking} needs, as cross-correlation, an external source. Overall, these three techniques do not lift the uncertainty on the spectral phase, which is a crucial piece of information for a complete electric field reconstruction required by the users, and for a better understanding of FEL gain dynamics.
Spectral phase interferometry for direct electric field reconstruction (SPIDER) has the potential for obtaining accurate and complete spectro-temporal information \cite{Walmsley} on a single shot by means of an inexpensive setup. The technique consists in measuring the spectral interference between the pulse itself and a replica delayed in time and slightly shifted in frequency. A direct inversion procedure of the interferogram yields the electric field of the pulse.

Though the implementation of a classical SPIDER apparatus remains rather simple at infrared, visible or even deep-ultraviolet wavelengths (at which it has been used on FEL pulses \cite{DUVFEL}), the issue of creating a replica at shorter wavelengths leads us to reconsider its arrangement, as was done on high order harmonic generation sources \cite{Mairesse, Cormier}.
Following the successful pump-probe experiments on the FERMI FEL facility \cite{NatComm} that were based on a double-seeding scheme, we here show that the same setup is suitable for carrying out SPIDER measurements. The arrangement we propose does not require any further implementation and is totally non-invasive for users since the on-line spectrometer \cite{spectro} can record the SPIDER interferogram.
The layout is shown in Fig. \ref{FEL}. 
Taking advantadge of the HGHG configuration, the requirement of a spectrally-sheared and temporally delayed replica is transferred to the seed before harmonic frequency upconversion to the XUV.
The seed is split in two replicas, temporally delayed and spectrally shifted. The successive seed pulses then interact with a sufficiently long and uniform electron bunch, so as to generate two spectrally sheared and temporally delayed FEL pulses. The two replicas of the measured pulse are thus directly produced, while in a classical SPIDER apparatus the pulse under study is indirectly characterized.

\begin{figure}[htbp]
\centerline{\includegraphics[width=1\columnwidth]{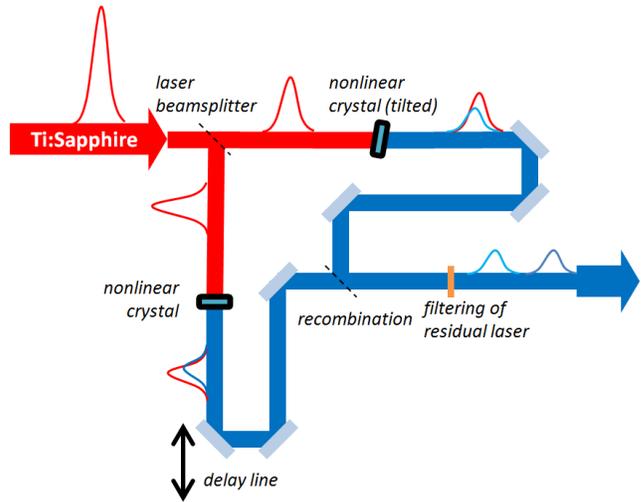}}
\caption{Test setup for double seed generation. The fundamental laser beam coming from a classical Ti:Sapphire femtosecond source is split in two paths. In each of them, an harmonic is generated in a nonlinear crystal. By slightly tilting a crystal, phase-matching conditions are changed, leading to a small and controllable spectral shift of the generated harmonic. The two beams are then recombined, a micrometric-precision motorized stage allowing to adjust the delay between the two pulses. A filter is eventually placed after recombination to leave out the residual part of the fundamental laser.}
\label{seed}
\end{figure}

\begin{figure}
\centering
\begin{tabular}{c}
\includegraphics[width=0.9\linewidth]{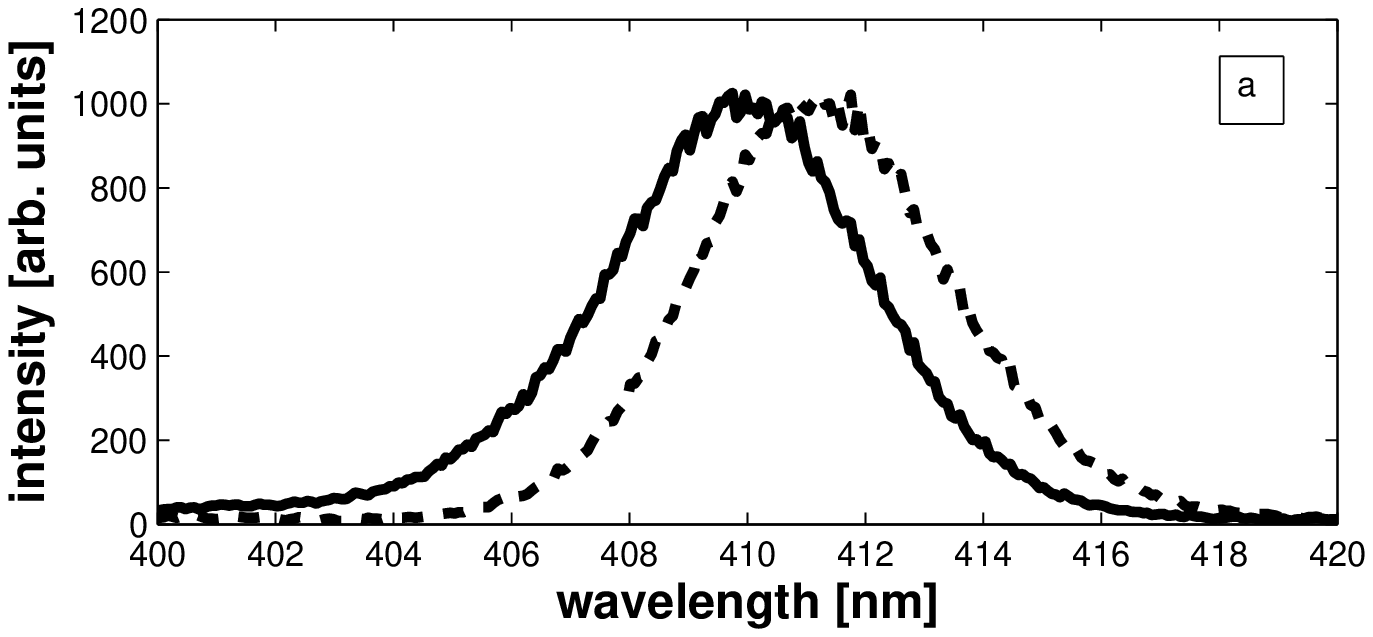} \\
\includegraphics[width=0.9\linewidth]{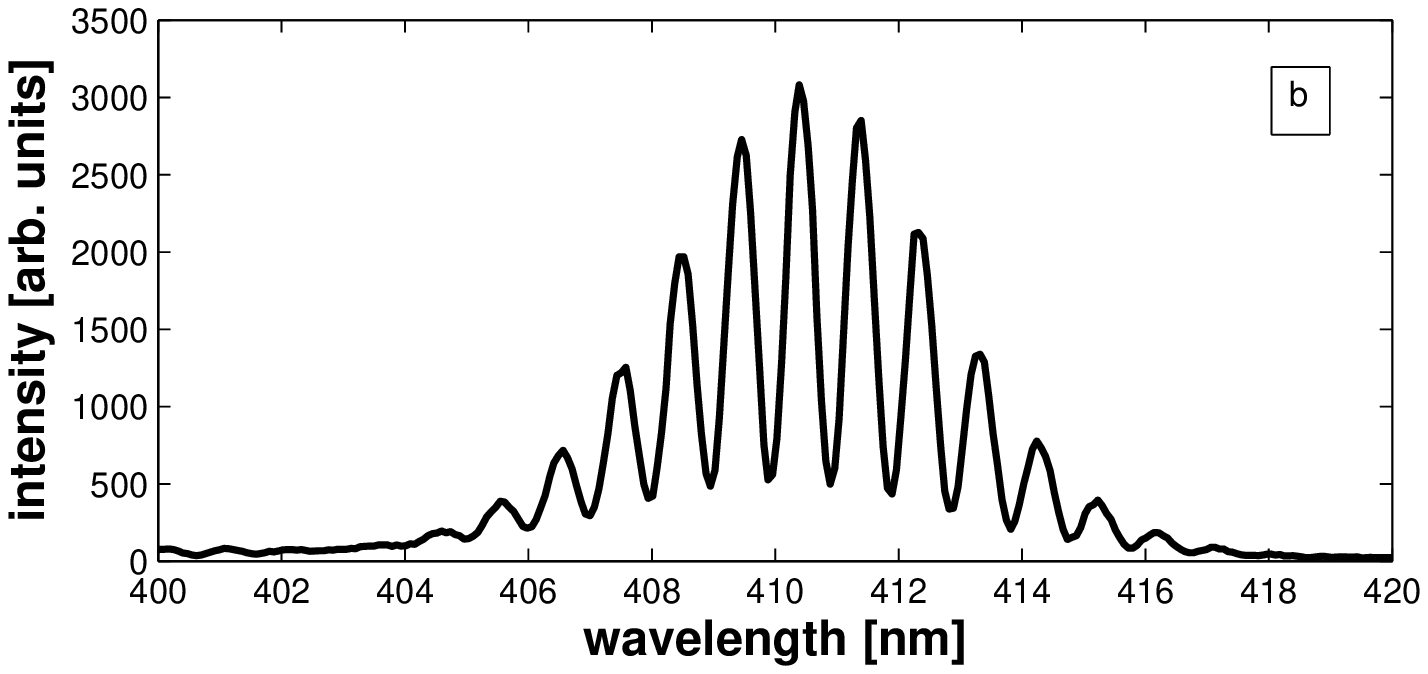} \\
\includegraphics[width=0.9\linewidth]{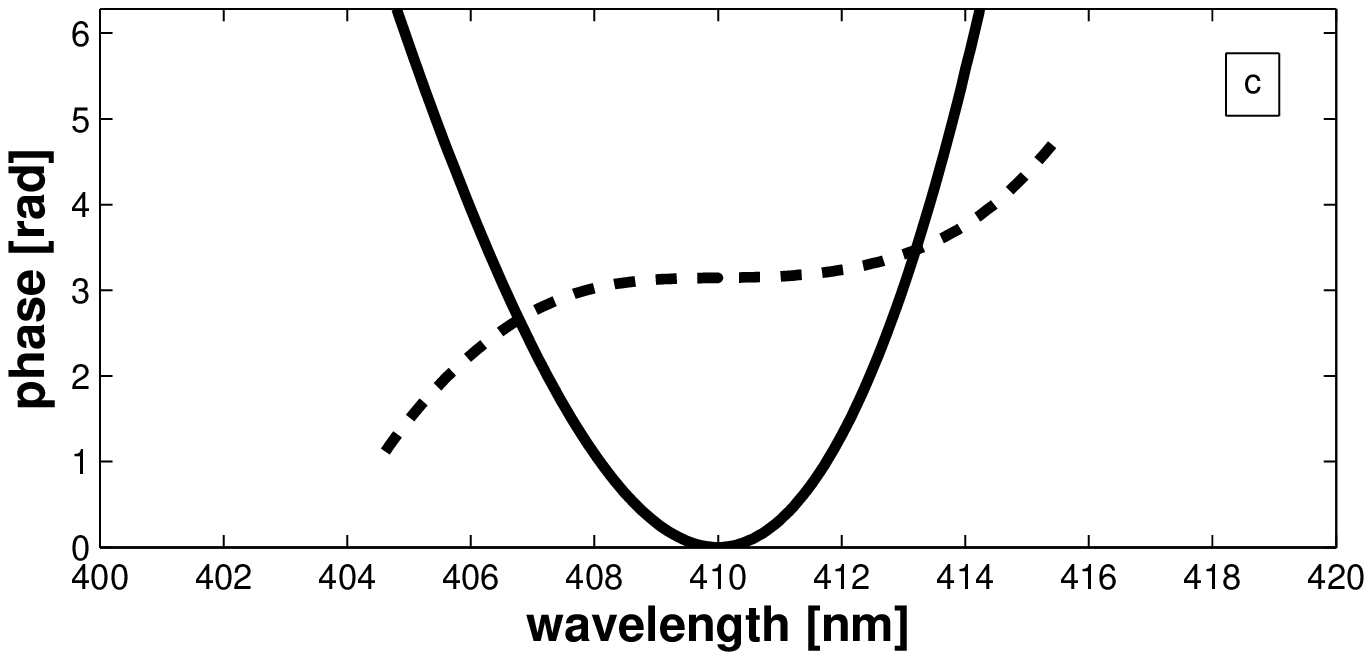}
\end{tabular}
\caption{Experimental characterization of the Mach-Zehnder setup. (a) measured reference spectrum (plain line) and spectrally sheared replica (dashed line, intensity multiplied by a factor 1.5). (b) measured interferogram, with a delay of $600~fs$ between the two replicas. (c) SPIDER retrieval of the phase with (plain line) and without (dashed line) a $1~cm$--thick fused silica plate placed after recombination. Single spectra and interferogram correspond to the integration of 3 successive pulses.}
\label{SPIDER_seed}
\end{figure}

As a first step, we wish to show the possibility of generating conveniently two similar seed pulses.
An experimental demonstration has been carried out at the Laboratoire d'Optique Appliqu\'{e}e, on a Ti:Sapphire source delivering $5~mJ$ pulses of duration $45~fs$, centered at a wavelength of $815~nm$, at a repetition rate of $1~kHz$ using a Mach-Zehnder setup depicted in Fig. \ref{seed}. On each arm of the interferometer, a $200~\mu m$--thick BBO type I crystal in the \textit{ooe} configuration has been chosen, thus generating the second harmonic of the fundamental beam, with an efficiency of about $15\%$. 
The advantadge of a Mach-Zehnder configuration is to independently control beam properties on each arm so that, as shown in Fig. \ref{SPIDER_seed}a, two spectrally sheared replicas are easily produced. Incidentally, deeper control, like variable attenuation or spatial shaping, can eventually be done on each arm for further applications.

Before extending our study to the XUV, a SPIDER experimental analysis on the seeding stage under test has been performed.
When the two replicas are sent together, with an appropriate delay, to the visible spectrometer, an interferogram is observed, such as shown in Fig. \ref{SPIDER_seed}b. This trace can then be analyzed by the algorithm described in \cite{Iaconis}, resulting in the calculation of the spectral phase (Fig. \ref{SPIDER_seed}c). When no dispersive element is present, a small residual third order dispersion is retrieved (dashed line), probably ensuing from the compressor settings. As expected, quadratic phase shows up when adding dispersive material on the beam path (plain line).

For the FEL simulations, we chose to consider realistic FERMI FEL parameters (see \cite{NaturePhotFERMI,NatComm,OptExpBM}), summarized in Table \ref{table}.
The seeding wavelength used here, corresponding to the third harmonic of a Ti:Sapphire source, enables to adopt a similar double-seeding scheme as described above.
\begin{table}[h!]
  \caption{Main parameters used for simulations. FWHM stands for the full-width at half-maximum value of the intensity. Undulator characteristics can be found in \cite{FERMICDR}.}
  \begin{center}
    \begin{tabular}{cc}
    \hline
    \textit{electron beam} & \\
    \hline
    mean energy & $1.2~GeV$ \\
    energy profile & flat \\
    current profile & flat \\
    peak current & $500~A$ \\
    duration & $1~ps$ \\
    \hline
    \textit{seed pulse} & \\
    \hline
    central wavelength & $261~nm$\\
    temporal profile & Gaussian \\
    bandwidth FWHM $\Delta\omega_0$ & $4.3 \cdot 10^{15}~rad\cdot s^{-1}$ \\
    energy & $10 - 100~\mu J$ \\
    \hline
    \textit{others} & \\
    \hline
    harmonic number $n$ & $6$\\
    FEL bandwidth FWHM $\Delta\omega_n$ & $\sim 1 \cdot 10^{14}~rad\cdot s^{-1}$ \\
    spectrometer resolution $d\omega_n$ & $2.13 \cdot 10^{12}~rad\cdot s^{-1}$ \\
    \hline
    \end{tabular}
  \end{center}
\label{table}
\end{table}

We need to choose the delay $\tau$ between the two pulses according to two conditions. First, in order to have a correct fringe resolution on the spectrometer detector, $\tau$ must be smaller than $\pi / d\omega_n = 1.5~ps$. Also, to obtain enough fringes, $\tau$ must be higher than $4\pi / \Delta\omega_n \approx 100~fs$. We thus choose $\tau=600~fs$. Then, the spectral shear between the interfering FEL pulses, equal to $n\Omega$, has to be choosen so as to satisfy the Whittaker-Shannon criterion, that is $n\Omega<2\pi/T$, where $T$ is the time interval whithin which the FEL pulse is expected to be contained. A value of $T=200~fs$ seems reasonable, which gives $n\Omega \approx 5 \cdot 10^{12}~rad \cdot s^{-1}$. This corresponds, on the seed pulses, to a shear of $0.2~nm$ in terms of wavelength. We opt for a value of $0.1~nm$ in the simulations i.e., $10\%$ of the bandwidth.
The FEL simulations have been carried out with the Perseo code \cite{Luca}. The simulation output provides the temporal profile of the FEL emission and, by Fourier transform, the FEL spectrum that is the SPIDER interferogram. Besides, noise is artificially added to the interferogram and a background threshold is taken into account, so as to come closer to the experiment.
Furthermore, in order to consider possible uncertainties and inhomogeneities (seed and current shape, phase and bandwidth of the seed pulses, slice emittance and energy spread along the electron beam, relative intensity of FEL pulses, etc.), the output phase and amplitude have been independently multiplied by a random error included within a range of $\pm 5\%$ of the calculated value.

\begin{figure}[htbp]
\centerline{\includegraphics[width=1\columnwidth]{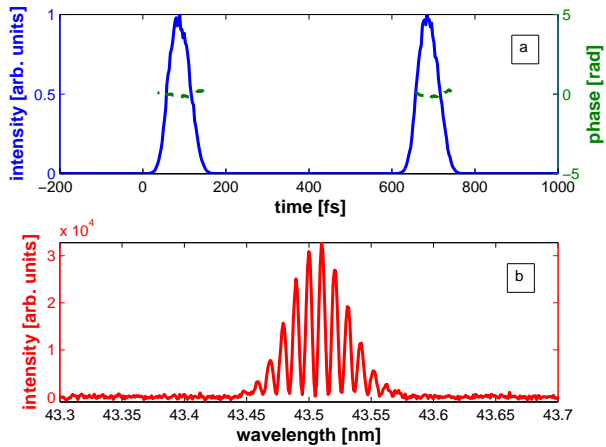}}
\caption{FEL simulation output with Fourier-transform limited seed pulses. (a) Longitudinal profile (plain line) and phase (line). (b) Spectrum.}
\label{simusFEL}
\end{figure}

\begin{figure}[htbp]
\centerline{\includegraphics[width=1\columnwidth]{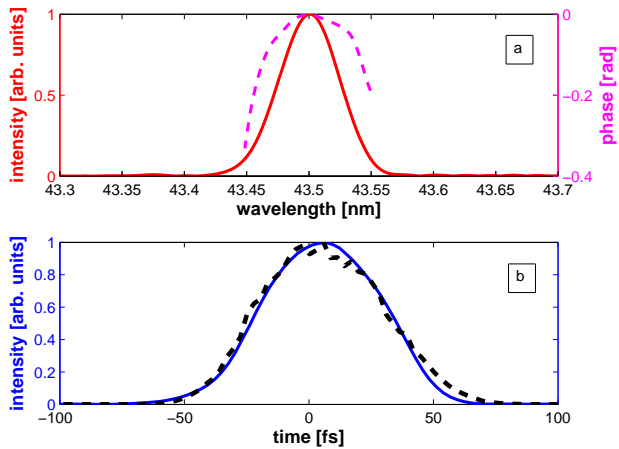}}
\caption{SPIDER reconstruction corresponding to FEL simulations carried out with Fourier-transform limited seed. (a) Spectrum (plain line) and spectral phase (dashed line). (b) Temporal intensity (plain line) compared to the direct output of the FEL simulation (dashed line).}
\label{SPIDER_FTL}
\end{figure}

Seed pulses with flat phase are first considered. The output of the simulation is shown in Fig. \ref{simusFEL}. As expected, the shapes of the FEL pulses mirror the one of the seed i.e., a Gaussian profile (Fig. \ref{simusFEL}a). Both are $60fs$-long (measured at FWHM) and their phases have very low distortions. The spectral interference pattern (Fig. \ref{simusFEL}b) of the FEL emission presents about 10 fringes within a wavelength window of $0.1~nm$. According to the FERMI spectrometer resolution at $43.5~nm$ (see Table \ref{table}), a maximum number of about 15 fringes can be resolved within this spectral window.
The temporal intensity retrieved by the SPIDER calculation is shown in Fig. \ref{SPIDER_FTL}, where it is overlapped with the first pulse (towards the head of the bunch and negative times in Fig. \ref{simusFEL}a) of the FEL simulation, taken as the reference one. As it can be seen, the agreement is very satisfactory and the FWHM duration obtained from the SPIDER reconstruction is only slightly shorter than the FEL simulation ($55~fs$).

\begin{figure}[htbp]
\centerline{\includegraphics[width=1\columnwidth]{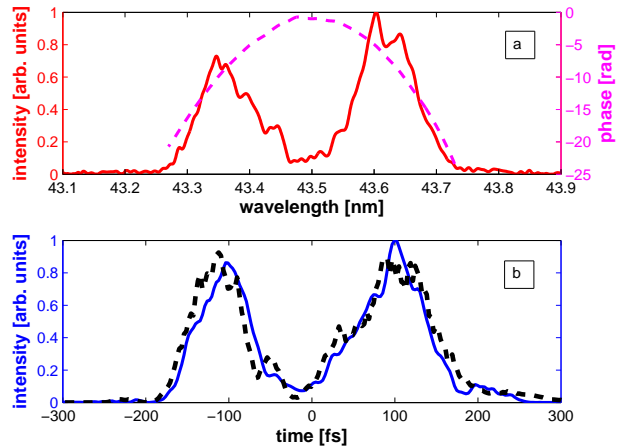}}
\caption{SPIDER reconstruction corresponding to FEL simulations carried out in double-peak regime. (a) Spectrum (plain line) and spectral phase (dashed line). (b) Temporal intensity (plain line) compared to the direct output of the FEL simulation (dashed line).}
\label{SPIDER_DP}
\end{figure}

In order to better test the robustness of the technique, we investigate thereafter a particular FEL regime. As detailed in \cite{PRLGDN,OptExpBM}, in specific conditions (intense seed and/or strong dispersive section, plus chirped seed), the FEL emission temporally splits into two sub-pulses, each with a distinct central wavelength. This two-colour structure can be seen in the SPIDER reconstruction of the spectrum (Fig. \ref{SPIDER_DP}a). The curve in the temporal domain (Fig. \ref{SPIDER_DP}b) shows that the reconstruction remains quite good: the split structure is correctly recovered and the duration of each sub-pulse is close to the FEL simulation as is the duration of the overall envelope ($\approx 300~fs$).
Notably, quite successful SPIDER reconstructions, not shown here, have also been obtained for different kinds of distorted FEL emission, resulting from other specific sets of machine parameters.

\begin{figure}[htbp]
\centerline{\includegraphics[width=0.8\columnwidth]{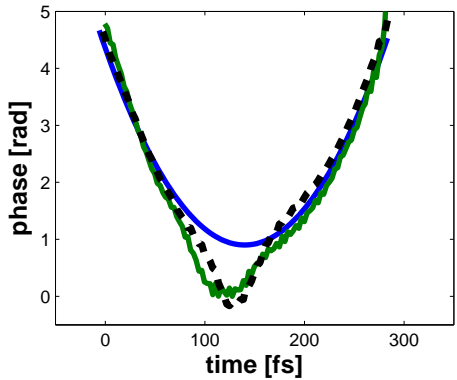}}
\caption{Calculated distorted FEL phase (dashed line) and its SPIDER reconstruction (green plain line). The parabola in blue plain line represents the direct phase transfer from the seed, calculated as the seed phase times the harmonic order.}
\label{phase_distortions}
\end{figure}

Let us now stress the relevance of FEL phase measurement. The theory \cite{Stupakov,Ratner} predicts that the seed phase structure is transferred to the FEL emission. It can nonetheless be distorted depending on the amplification regime. This is clearly observable in Fig. \ref{phase_distortions}, where the FEL emission is associated to a strong dispersive section. The parabola in blue plain line represents the direct phase transfer from the seed to the FEL emission. At positions where there is no saturation (before $70~fs$ and after $170~fs$, approximately), the FEL phase fits well the parabola.
Slight differences can be observed between the FEL phase output and its SPIDER retrieval, due to small dissimilarities of the replicas. They have two main causes: first, the uncertainties that we intentionally inserted within the numerical process; second, longitudinal distortions occurring in deep saturation. Very deep saturations may thus be a limitation of our technique.
Nevertheless, this study underlines the attractiveness of FEL SPIDER measurements for investigating the longitudinal gain dynamics in the FEL and the consequences of errors or high-order terms in the seed phase and of electron beam structurations. More complex regimes, including stronger phase distortions, can then be approached.
Such investigations could be experimentally accomplished by performing a SPIDER characterization on the FEL emission and, in parallel, on the seed (see Fig. \ref{SPIDER_seed}), building therewith a control loop for efficient FEL pulse shaping.

To conclude, the feasibility of a SPIDER experiment on a seeded FEL has been demonstrated, especially in the attractive case of XUV radiation produced in a HGHG scheme. The latter does not represent a limit: our scheme can be expanded, for instance, to direct seeding with high order harmonics \cite{seedingHHG}, echo-enabled harmonic generation \cite{EEHG} or self-seeding \cite{selfseeding,Lutman2}.
The main requirement of the proposed scheme is a sufficently long and uniform electron bunch, which has already been experimentally achieved. The method is then non-invasive and single-shot, can be performed in real time (together with experiments), and has the great advantadge to pull the necessary handling onto the seed instead of electron beam or XUV field. Besides, control of the FEL chirp by the seed makes feasible the generation of Fourier-limited FEL pulses that can then be measured by SPIDER.

We also demonstrated the large scope and the robustness of this method by studying a peculiar case of non-Gaussian, double-peak, emission. The characterization of the latter has a significant interest since such an emission is suitable for XUV pump-probe experiments. Moreover, a single-seed SPIDER experiment in double-peak regime could be foreseen: the split sub-pulses, originating from one seed only, may be two suitable replicas (see Fig. \ref{SPIDER_DP}).

Finally, experimental demonstration of the XUV FEL SPIDER would allow to measure the effective transfer of the seed phase and possible phase distortions, occurring through FEL amplification in different regimes. This will have an important impact for a better understanding of FEL physics and for further developments, such as the design of chirped-pulse amplification schemes \cite{CPAYu,CPAFrassetto,CPAFeng,CPAOliva}.

We gratefully acknowledge the COST MP1203 action for its financial support and the FERMI team for constructive discussions.

\end{document}